\documentclass[11pt]{article} 
\usepackage[extreme]{savetrees}

\usepackage[utf8]{inputenc} 

\usepackage{geometry} 
\usepackage{graphicx} 

\usepackage{booktabs} 
\usepackage{array} 
\usepackage{paralist} 
\usepackage{verbatim} 
\usepackage{subfig} 

\usepackage{fancyhdr} 
\usepackage{sectsty}
\allsectionsfont{\sffamily\mdseries\upshape} 

\usepackage[nottoc,notlof,notlot]{tocbibind} 
\usepackage[titles,subfigure]{tocloft} 

\title{A Tale of Tail Covariances (and Diversified Tails)}
\author{Jan Rosenzweig}
\date{} 

\begin{document}
\maketitle

\begin{abstract}
This paper deals with tail diversification in financial time series through the concept of statistical independence by way of differential entropy and mutual information.
By using moments as contrast functions to isolate the tails of the return distributions, we recover the {\it tail covariance matrix}, a specific two-dimensional slice of the mixed moment tensor, as a key driver of tail diversification. 

We further explore the links between the moment contrast approach and the original entropy formulation, and show an example of in- and out-of-sample diversification on a broad stock universe.
\end{abstract}

\section{Introduction}

Diversification of portfolios in the presence of fat-tailed returns is an ongoing problem in finance, with many open questions still unresolved. 

At the crux of the problem is the fact that our go-to measure of (in)dependence of returns, correlation, only actually measures independence if the underlying returns are Gaussian, and coupled via a Gaussian copula. If the returns distributions are not Gaussian, or if they are coupled through a non-Gaussian copula, then zero correlation does not imply statistical independence -- a fact which even has a Wikipedia page dedicated to it \cite{wiki}.

In the standard Gaussian framework, this is usually handled using some combination of local volatility, stochastic volatility, local correlation and stochastic correlation. But this approach often boils back down  to diversification through decorrelation, which delivers, at best, mixed results.

An additional difficulty lies in the fact that extreme returns, while being more frequent than the Gaussian distribution would imply, are still comparatively rare; and financial time series are non-stationary, so any relevance of historical observations decays reasonably quickly with time.

A number of approaches exist to handle those issues, reviewed on a high level in \cite{bouchaud}.

We take a somewhat different approach, directly based on the notion of statistical independence, specifically through mutual information and the independent component analysis. This follows on some earlier work that looked at similar approaches in the context of portfolio optimization and hedging \cite{FTF, PLP, LDP}, but not directly in the context of diversification.

The paper is organised as follows: Section 2 reviews the use of moments to isolate the tails of returns distributions; Section 3 sets out the mutual information approach and the ICA methods; Section 4 looks at tail-focused ICA using the moment contrast function; Section 5 integrates the moment contrast approach with the original entropy formulation; Section 6 describes an experiment on Russell 3000 stocks; and Section 7 summarizes the conclusions.

\section{Moments and Tails}

Useful information on the tails of a distribution can be gleaned from its moments of high order \cite{PLP}; for a finite sample $x_{1},x_{2},...,x_{m}$ of a centered random variable $x$, its moment of order $p>0$ is
\begin{equation}
    M_{p}(x) = \frac{1}{m} \sum_{i=1}^{m}   x_{i}^{p} =  \frac{1}{m} x_{\infty}^{p}\sum_{i=1}^{m} \left( \frac{ x_{i}}{x_{\infty}} \right)^{p}
    \label{moment}
\end{equation}
where 
$$x_{\infty} = \max_{i} |x_{i}|.$$

\begin{figure}[!htb]
	\includegraphics[width=0.48\textwidth]{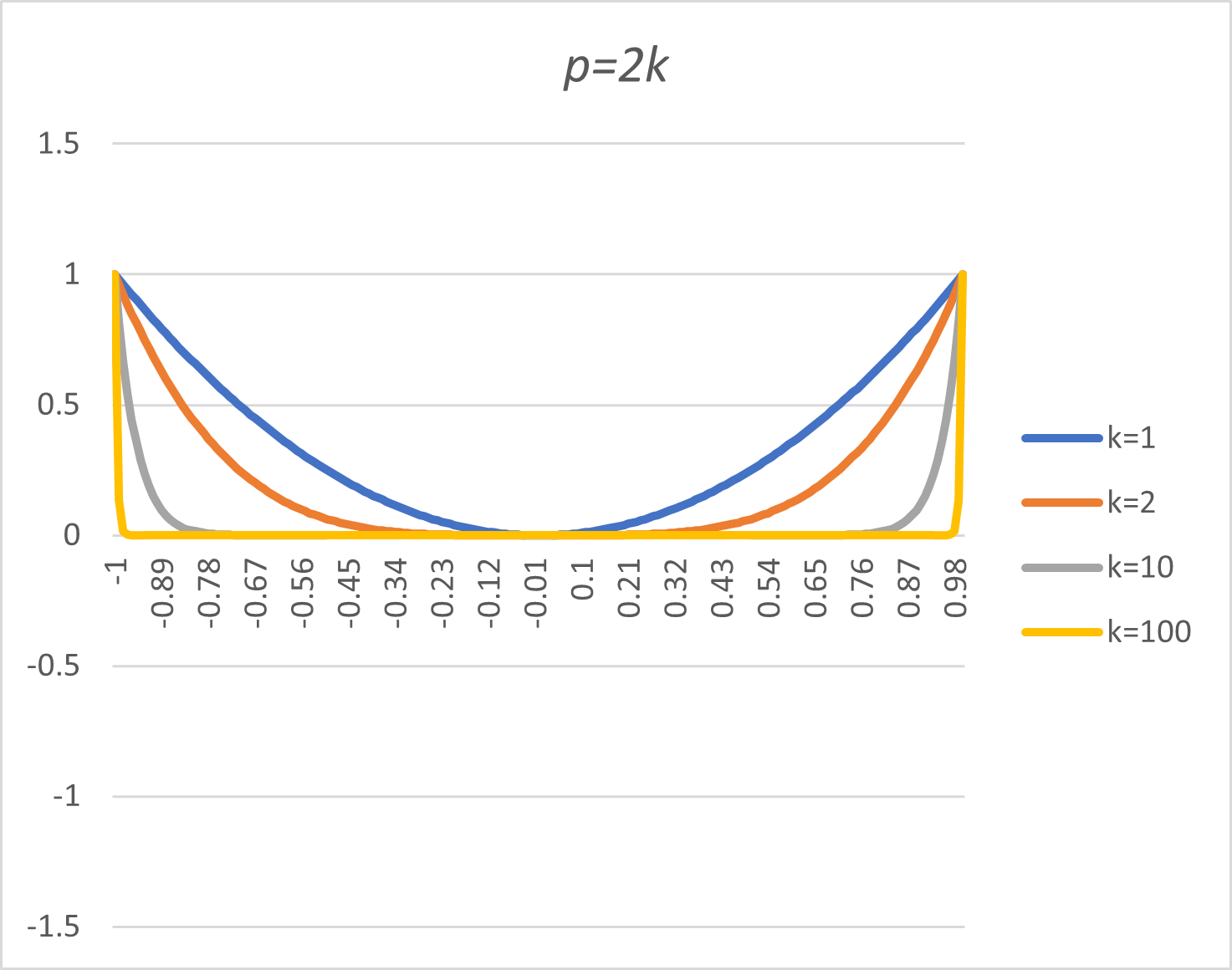}
	\includegraphics[width=0.48\textwidth]{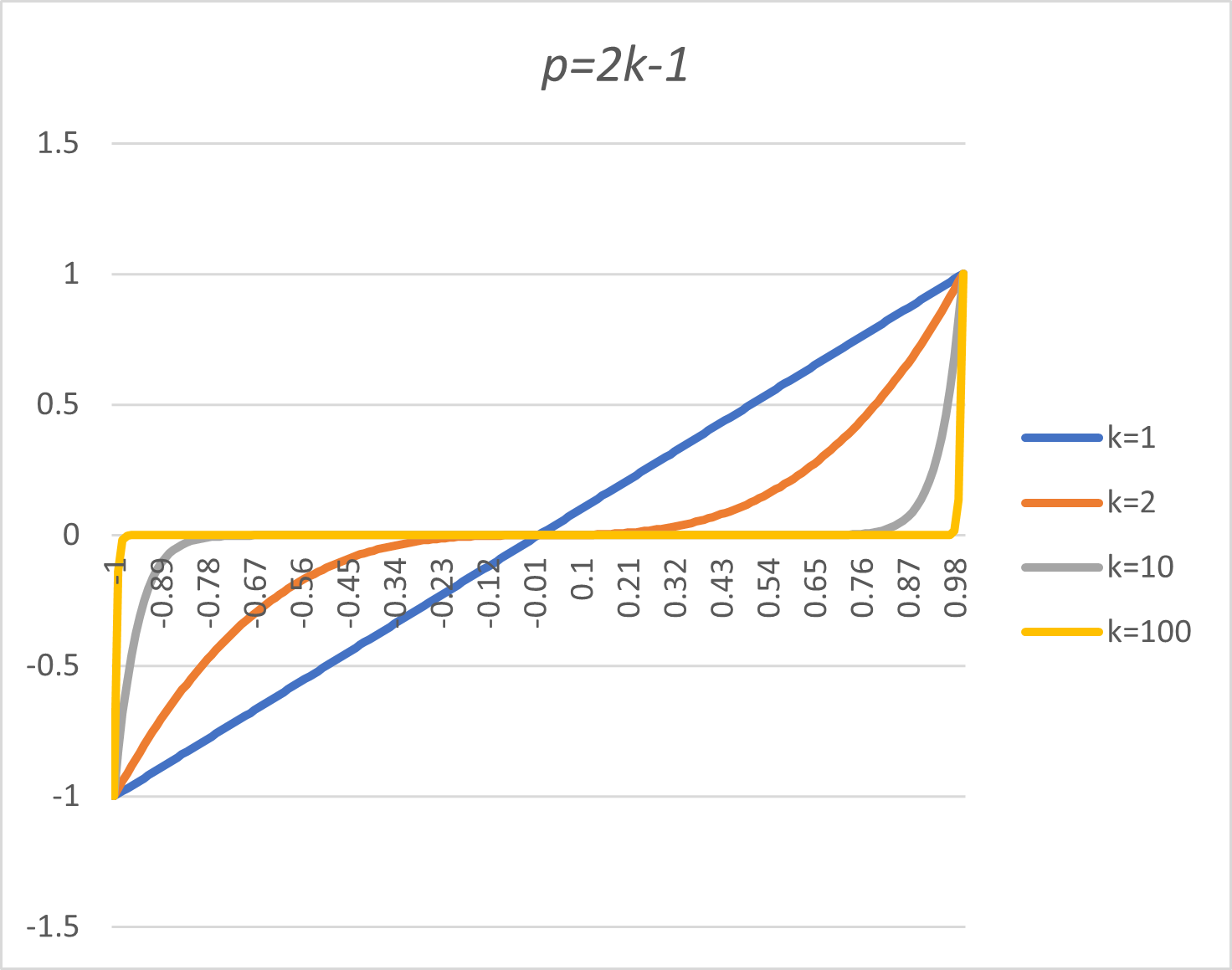}
	\caption{Moments of even and odd orders. }
	\label{fig:moment}
\end{figure}

Since each ratio $x_{i}/x_{\infty}$ is within $-1 \le x_{i}/x_{\infty}\le 1$, in the limit of $p \nearrow \infty$, each $(x_{i}/x_{\infty})^{p}$ tends to either $\pm 1$, or zero. In other words, moments effectively act as high pass amplitude filters, in the sense that elements from the centre of the distribution are comparatively suppressed, and only elements from the tail of the distribution contribute to the moment. The effective width of the band of contributing elements decays to zero as $p \nearrow \infty$, and, in that limit, only the largest element by absolute value contributes to the moment.
For moments of even order, $p=2k$, this reduces to the familiar result from calculus
$$\lim_{k\rightarrow \infty} M_{2k}^{1/2k} = x_{\infty},$$
and, in terms of
$$x_{max} = \max_{i} \ x_{i},\ \ x_{min} = \min_{i}\ x_{i},$$
we can more generally write
$$
\lim_{k\rightarrow \infty} M_{2k}^{1/2k} = \max( x_{max}, - x_{min}), \
$$
\begin{equation}
\lim_{k\rightarrow \infty} M_{2k-1}^{1/(2k-1)} = \left\{ \begin{array}{lcl}
x_{max} & | & x_{max} > -x_{min}\\ 
x_{min} & | &  x_{max} < -x_{min}.
 \end{array} \right.
\label{momentex}
\end{equation}
Note that there are multiple possibilities for $M_{2k-1}^{1/(2k-1)}$ when $x_{min}+x_{max}=0$, depending on how many points reach either of $x_{min}$ or $x_{max}$ -- the result could be $x_{min}$, $x_{max}$ or $0$. When the underlying distribution is continuous, however, probability of repeated values is zero, and hence we can disregard this case.

The advantage of using moments to isolate tails of the distribution is that moments are smooth functions of sample elements, while $\max$ and $\min$ are not; hence, by using moments, we open the door to a number of numerical methods that would be unavailable when using $\max / \min$, ranging from simple Newton's iteration, to more complex methods.

\section{Entropy, Independence and Mutual Information}

By definition, a random vector ${\bf y}=(y_{i})$ with a probability density function $\rho({\bf y})$ is independent if the density is the product of its marginals,
\begin{equation}
\rho({\bf y}) = \prod_{i} \rho_{i}(y_{i}), \label{independence}
\end{equation}
hence any measure of functional distance between $\rho({\bf y})$ and $\prod_{i} \rho_{i}(y_{i})$ is a measure of mutual dependence between $y_{i}$.

Significant advances for continuous random variables have been achieved by using the concept of differential entropy, defined as 
\begin{equation}
H({\bf y}) = - \int \rho({\bf y}) \ln \rho({\bf y}) d{\bf y},
\label{entropy}
\end{equation}
and the {\it Mutual Information } of ${\bf y}$, defined as the Kuhlback-Lebler divergence between $\rho$ and $\prod \rho_{i}$,
$$I({\bf y}) =  \int \rho({\bf y}) \ln \frac{\prod_{i}\rho_{i}(y_{i})}{ \rho({\bf y})} d{\bf y} =  - \int \rho({\bf y}) \ln \rho({\bf y})d{\bf y} + \sum_{i=1}^{n} \int \rho({\bf y}) \ln \rho_{i}(y_{i})d{\bf y}  =$$
\begin{equation}
=H({\bf y}) - \sum_{i=1}^{n} H_{i}(y_{i}), \label{mi}
\end{equation}
where
$$H_{i}(y_{i}) = - \int \rho({\bf y}) \ln \rho_{i}(y_{i})d{\bf y} = - \int \rho_{i}( y_{i}) \ln \rho_{i}(y_{i})d{\bf y}$$
is the marginal entropy of $y_{i}$. This has the straightforward intepretation that the mutual information is all the information contributed by  ${\bf y}$ jointly, minus the information contributed by each of the $y_{i}$ individually.

This has lead to the development of a range of methods for decomposing random vectors into maximally mutually independent components by minimizing their mutual information, collectively referred to as Independent Component Analysis (ICA) \cite{ica}. Briefly, the key insight is that any decomposition is just a change of coordinates, which does not affect the joint entropy term. Therefore, one only has to pick orthogonal directions which minimize the marginal entropies. A further piece of insight is that, for given variance, entropy is maximized by the Gaussian distribution. Hence,  directions of minimal marginal entropy are also  the directions of maximal non-Gaussianity, using that particular measure of non-Gaussianity.

We can then approximate the marginal entropies in the form \cite{mutual} 
$$H_{i}(y_{i}) \approx E (G(y_{i}))$$
for some smooth, non-quadratic scalar contrast function $G$, typically chosen to isolate a relevant measure of non-Gaussianity, and the maximal mutual independence reduces to solving $n$ parallel constrained maximization problems with coupled constraints,
\begin{equation}
E (G({\bf w}_{i}^{\tau} {\bf y})) \rightarrow \max, \ \ E({\bf W}^{\tau}{\bf y})^{2} = {\bf 1}.
\label{opt}
\end{equation}
where ${\bf W}$ is the unknown  $n \times n$ weights matrix with columns ${\bf w}_{i}$, and ${\bf 1}$ is the $n \times n$ unit matrix. The Karush-Kuhn-Tucker conditions can be reduced to, in vector form,
\begin{equation}
E\left( {\bf W}^{\tau} {\bf y} g( {\bf W}^{\tau}{\bf y}) - diag( {\bf w}_{i}^{\tau}{\bf y} g( {\bf w}_{i}^{\tau}{\bf y}) ) \right) = {\bf0}, \ {\bf W}^{\tau} {\bf W} = {\bf 1}
\label{kkt}
\end{equation}
where $g = G'$, and it is applied component-wise to the vector ${\bf W}^{\tau}{\bf y} = ({\bf w}_{i}^{\tau}{\bf y})_{i}$ \cite{mutual}. This can then be solved numerically by using a fast semi-Newton iteration \cite{fastica}.

We note that the system (\ref{kkt}) is overdetermined, since it imposes more constraints than there are tunable parameters. The number of recoverable independent sources is therefore generally less than the size of the underlying universe available for portfolio construction.

\section{Isolating the Tails -- Moments as Contrast Functions}

In the context of tail diversification, the task is to pick a contrast function for equation (\ref{opt}) in such a way that the resulting components retain their diversification in the tails of the distribution. In view of the tail-focusing properties of moments as discussed in Section 2, we use
\begin{equation}
G(u) = \frac{1}{2k} u^{2k},\ g(u) = u^{2k-1}.
\label{contrast}
\end{equation}
Then, equation (\ref{kkt}) becomes
\begin{equation}
\begin{array}{llr}
E( {\bf w}_{i}^{\tau} {\bf y} ( {\bf w}_{j}^{\tau}{\bf y})^{2k-1}) = { 0}, & {\bf w}_{i}^{\tau} {\bf w}_{j} = 0 & | \ i\neq j\\
{\bf w}_{i}^{\tau} {\bf w}_{i} = 1.& &
\end{array}
\end{equation}

We call the matrix
\begin{equation}
{\bf T}^{(k)} = \left( E( {\bf w}_{i}^{\tau} {\bf y} ( {\bf w}_{j}^{\tau}{\bf y})^{2k-1}) \right)_{ij}
\end{equation}
the {\it tail covariance matrix} of order $k$ for the vector ${\bf W}^{\tau} {\bf y}$. It is a straightforward generalization of the standard covariance matrix, which we recover by setting $k=1$. The tail covariance matrix is, however, generally not symmetric for $k>1$.

The interpretation of tail covariances is straightforward in terms of Figure \ref{fig:moment}; order $k$ tail covariance of $x$ and $y$ is the covariance of $x$ with the tails of $y$, filtered using the moment of order $2k-1$. This also provides an intuitive explanation for  why tail covariances are not symmetric; covariance of $x$ with the tails of $y$ is generally not the same as the covariance of $y$ with the tails of $x$. 

In the limit of $k \nearrow \infty$, the filter retains only the element with the highest absolute value, and the tail covariance of $x$ and $y$ is simply the overlap of $x$ with the extreme tail event of $y$.

Tail covariances were observed in the finance context in \cite{LDP}, in connection with hedging tail risk in liability-driven portfolios. Geometrically, they can be interpreted as projections in ${\mathcal L}_{p}$ norm, i.e. the the nearest point on a hyperplane of an $n$-dimensional space to the origin, in  ${\mathcal L}_{p}$-norm \cite{lp}.

We can now safely interpret in what sense the independence is optimized by using (\ref{opt}) with the moment contrast function (\ref{contrast}); they generate components that have linear covariances equal to zero, and order-$k$ tail covariances equal to zero.

This still leaves a number of other covariances of the same order, namely covariances of the form $E(x^{j}y^{2k-j})$ for general values of $j$.

In general, independence of centered random variables $x$ and $y$ only implies that their moments are separable;
$$E\left( x^{j} y^{2k-j} \right)= E\left( x^{j} \right) E\left( y^{2k-j} \right).$$
These only automatically equal zero when either $j=1$ or $2k-j=1$ due to the centeredness of the distributions, $E(x)=E(y)=0$. It is therefore somewhat welcome that general mixed moments do not appear in our derivation of independence; since they do not vanish for independent random variables, it is less straightforward to use them as a measure of independence. Coming back to the earlier point about the system already being overdetermined just by using tail covariances, this is not a bad thing - any additional restrictions on covariances of arbitrary order would only make it even more overdetermined.

\section{What about Entropy?}

The choice of contrast function $G(u) = u^{2k}/2k$ was in many ways external to the task of maximizing the independence of the underlying variables; if we start from the assumption that there is no function $G$ that would lead to an acceptable approximation to the marginal entropy, we might as well use one that serves our specific purpose for other reasons.

We can, however, try to estimate marginal entropy directly from the data. We again start from a finite sample  $x_{1},x_{2},...,x_{m}$  as in Section 2, and sort it in increasing order, so that  $x_{i_{1}}\le x_{i_{2}}\le...\le x_{i_{m}}$. If the underlying distribution is continuous, the probability of repeated terms in a finite sample is zero, and the inequalities are all strict with probability 1. In statistical terms, $x_{i_{1}},...,x_{i_{m}}$ are order statistics for the sample $x_{1},x_{2},...,x_{m}$.

One approach would be to use a Vasicek-like entropy estimation \cite{vasicek}. Vasicek's formulation rests on the unknown cumulative density of $F(x)$ such that $F'(x) = \rho(x)$; the integration switches from integration over the domain of $x$ to integration over the probability $q$, and the entropy is expressed as 
\begin{equation}
H(x) = \int_{0}^{1} \ln \frac{d}{dq}F^{-1}(q) dq.
\label{vasicek}
\end{equation}
Using the estimate 
$$F\left(x_{i_{j}} \right)=\frac{j}{m+1},\ \ F^{-1}\left( \frac{j}{m+1}\right) = x_{i_{j}}$$
for $j=1,..,m$, we get the order-$n$ Vasicek estimarte of entropy by taking the $n$'th order finite difference of $x_{i_{j}}$,
$$
H(x) \approx \frac{1}{m+1} \sum_{j=1}^{m} \ln  \left[ \frac{m+1}{2n} \left( x_{i_{j+n}} - x_{i_{j-n}} \right) \right]
$$
with boundary conditions $x_{i_{j}}=x_{i_{1}}$ for $j<1$, $x_{i_{j}}=x_{i_{m}}$ for $j>m$. Asymptotically for large $x_{\max}$ and $x_{\min}$, this gives
\begin{equation}
    H(x) \approx   \ln\ -x_{\min} + \ln\ x_{\max}  + \ln \frac{m+1}{2n} + O \left( 1  \right). \label{vasicektail}
\end{equation}
From equation (\ref{moment}), we have
\begin{equation}
\ln M_{2k} = \ln \frac{1}{m} + 2k\ \ln\ x_{\infty} + \ln \sum_{i=1}^{m} \left( \frac{x_{i}}{x_{\infty}} \right)^{2k}.
\label{logmoment}
\end{equation}
The sum in (\ref{logmoment}) tends to $1$ with probability $1$ for large $k$, hence (\ref{vasicektail}) and (\ref{logmoment}) combine to
\begin{equation}
H(x) \approx \frac{1}{2k} \ln\ M_{2k} +\frac{1}{2k} \ln\ m +  \ln \frac{m+1}{2n} + O \left( 1  \right).
\label{logentropy1}
\end{equation}

The problem with the Vasicek entropy is that it effectively loads $x_{\min}$ and $x_{\max}$ $n$ times, getting $\delta$ mass and infinite slope of $F$ at both points, which is not consistent with a continuous density function $\rho$. Correcting the boundary conditions in the Vasicek formulation leads to the Ebrahimi estimator \cite{ebrahimi}, which eliminates the $n$-fold repetition of $x_{\max}$ and $x_{\min}$. Given the log dependence of the entropy estimate on the tails, that only results in the addition of a term $(\ln\ n) / (m+1) $, and it does not affect the asymptotic estimate at leading order.

In either case, on finite samples from continuous distributions, marginal entropy  is, to leading order, logarithmic in the expectation of our moment contrast function $G(u) = u^{2k}/2k$.

The key here is that we are only dealing with finite samples of continuous distributions, so, with probability 1, no point occurs more than once in the sample. In a continuous limit or with discrete distributions, this is clearly not the case, and the approximation  (\ref{logentropy1})  does not hold.

\section{Diversifying Russell 3000 stocks}

For a practical example, we look at Russell 3000 stocks over a period of eight years, from September 2014 until October 2022. We used percentage daily returns, adjusted for dividends and splits, and filled zero daily returns on any stock-specific non-trading days.

The time series was divided into two non-overlapping buckets of equal length, covering October 2014- September 2018, and October 2018- September 2022.

\begin{figure}[!htb]
	\includegraphics[width=0.8\textwidth]{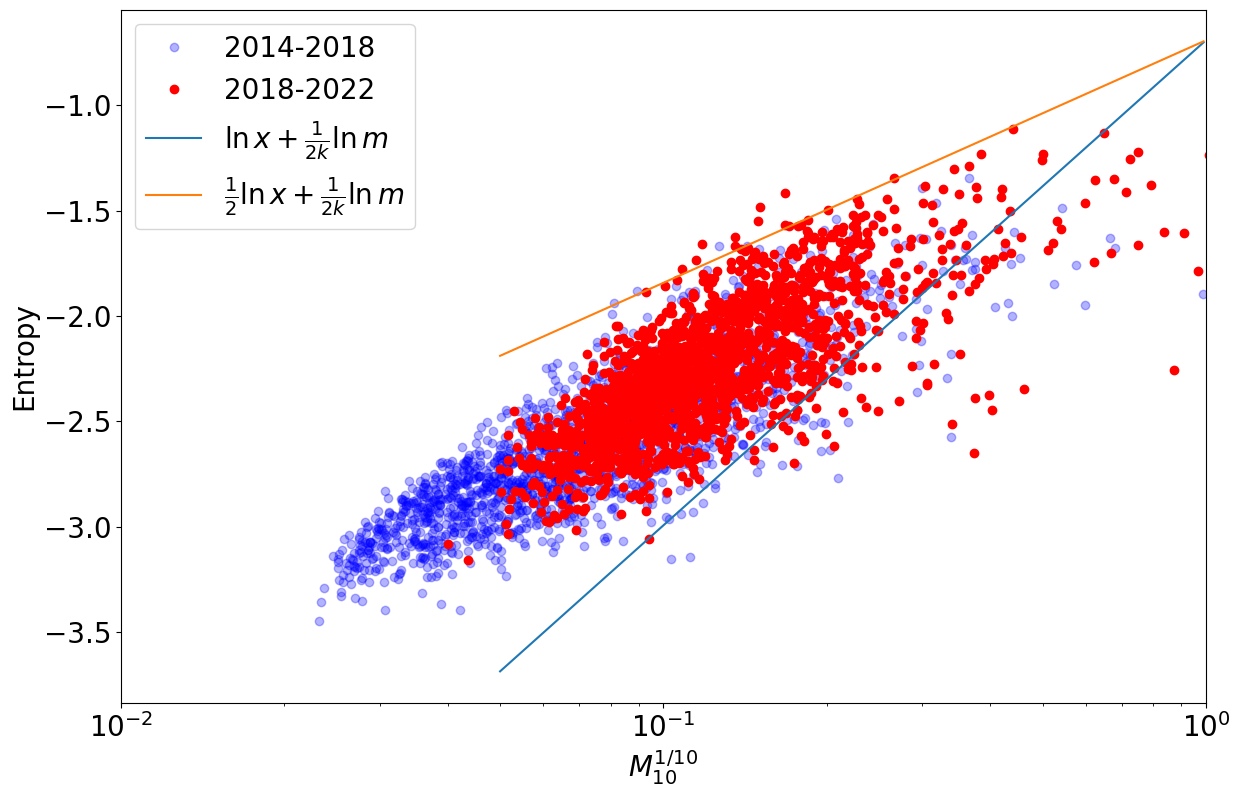}
	\caption{Russell 3000 stocks, $M_{10}^{1/10}$ of the daily returns on the $x$-axis, differential entropy on the $y$-axis. Each circle represents one symbol; blue circles for the 2014-2018 bucket, red circles for the 2018-2022 bucket. Entropy is on a linear scale, moment on a log scale.}
	\label{fig:entropy}
\end{figure}

We first plotted the relationship between the moments and differential entropy. To avoid biasing the entropy estimators, we used Correa's method for the entropy calculation. The results are shown in Figure \ref{fig:entropy}.

\begin{figure}[!htb]
\centering
	\includegraphics[width=0.95\textwidth]{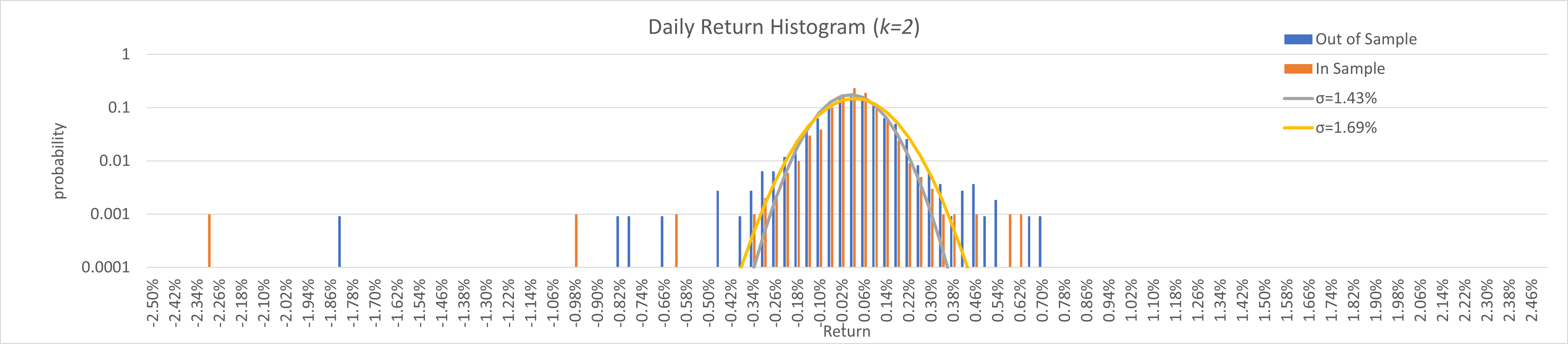}
 \centering
	\includegraphics[width=0.95\textwidth]{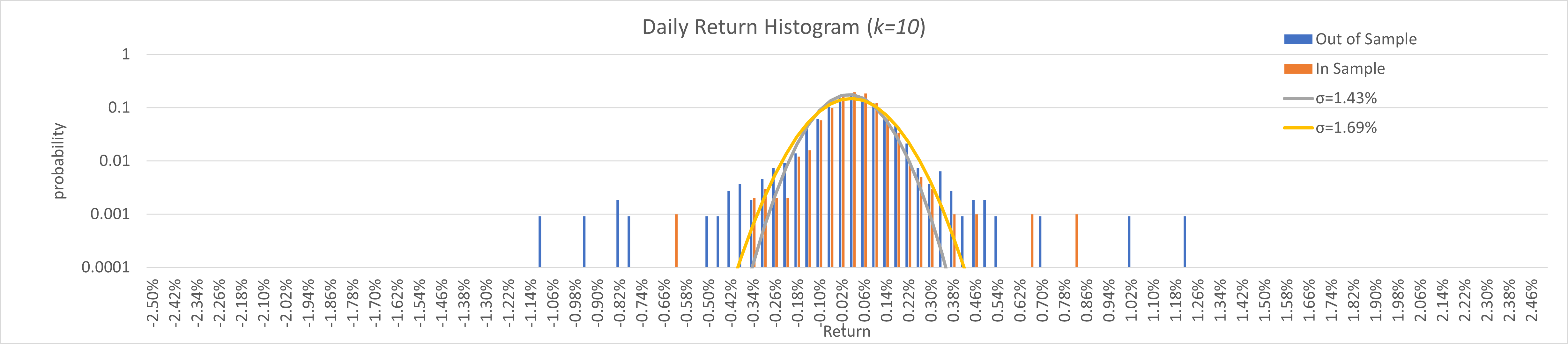}
	\caption{Russell 3000 stocks, equally weighted portfolios, histograms of in- and out-of-sample daily returns for $k=2$ and $k=10$.}
	\label{fig:result}
\end{figure}

As seen from Figure \ref{fig:entropy}, the logarithmic scaling of entropy with the moments is confirmed by the data. The correspondence is not $1-1$, due to the fact that there are significant contributions to entropy other than the moments; the association between entropy and log moment is, however,  nonetheless strongly positive, which further justifies the use of the moment contrast function as a stand-in for entropy.

We then proceeded to calibrate 500 independent components on the 2014-2018 bucket, and test their diversification properties on the 2018-2022 bucket. 

As we can see from Figure \ref{fig:result}, the out-of-sample tails are fairly consistent and controlled with increasing $k$. The curvature of the centre of the distribution, which can be taken to represent the Gaussian component of the underlying process, barely changes as we change the value of $k$. The tails, however, are clearly compressed towards the centre as a consequence of increasing $k$.
 
The cost of such compression of the tails with increasing penalty order $k$ is having more large returns of smaller magnitude. The underlying risk does not go away, but it is distributed closer to the centre of the distribution.

It is of note that the in-sample period 2014-2018 was fairly quiet, hence the relative rarity of tail events is understandable; the out-of-sample period 2018-2022 comprised both the Covid-19 induced crash in 2020, and the geopolitical events of 2021-2022, hence the relative increased frequency of tail events. Tail ICA nonetheless keeps a good handle on the tails in the volatile 2018-2022 out-of-sample bucket, and it compresses the tails progressively with increasing $k$.

\section{Conclusions}

The approach we presented in this paper is a new approach to constructing diversified components that retain their diversification in the tails of the distribution. 

Instead of contorting realistic distributions to fit into the mould of multivariate Gaussian distributions, we approach the problem from the angle of statistical independence, viewing the tail diversification problem as minimization of statistical dependence between diversified components. 

Our approach identifies {\it tail covariances}, or mixed moments of order $E(x y^{2k-1})$ and $E(x^{2k-1}y)$, as the key measures of tail independence of random variables $x$ and $y$, together with orthogonality. Tail covariances naturally come out of the mutual information minimization formulation when using a moment contrast function to isolate the tails; they can also be linked to the original entropy formulation of the mutual information minimization problem. 

In the example of Russell 3000 stocks, we have shown that, calibrating independent components on a relatively quiet period of 2014-2018, the resulting components were nonetheless able to retain diversification through the turbulent trading of 2020 and 2022.

\end{document}